# Unlocking Thermoelectric Potential: A Machine Learning Stacking Approach for Half Heusler Alloys

Vipin K E[1] and Prahallad Padhan[1,2]

[1]*Department of Physics, Nanoscale Physics Laboratory, Indian Institute of Technology Madras, Chennai 600036, India*
[2]*Functional Oxides Research Group, Indian Institute of Technology Madras, Chennai 600036, India*

**ABSTRACT**

Thermoelectric properties of Half Heusler alloys are predicted by adopting an ensemble modelling approach, specifically the stacking model integrated using Random Forest and XGBoost scheme. Leveraging a diverse dataset encompassing thermal conductivity, the Seebeck coefficient, electrical conductivity, and the figure of merit (ZT), the study demonstrates superior predictive performance of the stacking Model, outperforming individual base models with high $R^2$ values. Key features such as temperature, mean Covalent Radius, and average deviation of the Gibbs energy per atom emerge as critical influencers, highlighting their pivotal roles in optimizing thermoelectric behavior. The unification of Random Forest and XGBoost in the stacking model effectively captures nuanced relationships, offering a holistic understanding of thermoelectric performance in Half Heusler alloys. This work advances predictive modelling in thermoelectricity and provides valuable insights for strategic material design, paving the way for enhanced efficiency and performance in thermoelectric applications. The ensemble modelling framework, coupled with insightful feature selection and meticulous engineering, establishes a robust foundation for future research in pursuing high-performance thermoelectric materials.

**Introduction**

In pursuing sustainable energy solutions, thermoelectricity has emerged as a promising field, offering a unique mechanism to convert waste heat into electrical energy[1]. The foundation of thermoelectric devices lies in the intrinsic properties of thermoelectric materials. These materials possess the ability to generate an electric voltage when exposed to a temperature gradient and hold significant promise for applications in solid-state power generation and refrigeration, with notable uses in industrial dissipated heat recovery, solid-state refrigeration for electronic communications, and specialized deep-sea power supplies[2,3]. These materials offer noise-free and environmentally friendly alternatives, garnering global interest due to their straightforward structures and distinctive characteristics. The efficacy of thermoelectric materials is assessed through a dimensionless figure of merit (ZT), given by the formula:

$$ZT = \frac{(S^2 \sigma T)}{\kappa_l + \kappa_e}$$

where S is the Seebeck coefficient; T is the absolute temperature in Kelvin; σ is the electrical conductivity; and $k_L$ and $k_e$ represent the lattice thermal conductivity and electronic thermal conductivity, respectively. [4] Thermoelectric materials exhibiting superior performance are generally characterized by a higher power factor (PF), represented as $S^2\sigma$ coupled with lower thermal conductivity. Nevertheless, it is essential to acknowledge that the relationship between these parameters is intricately complex, and their modulation is not independent. Consequently, enhancing the ZT values of thermoelectric materials poses a significant challenge. [5] Among the various thermoelectric materials, half-Heusler (HH) compounds stand out as promising candidates for applications at mid- and high-temperatures[6]. Beyond their notable thermoelectric performance, these compounds boast favorable mechanical properties, thermal stability, and low toxicity, thereby holding considerable promise for practical applications[7,8]. HH compounds are ternary intermetallics characterized by a general formula XYZ, wherein X represents the most electropositive element, Y is a less electropositive transition metal, and Z is a p-block element. These compounds exhibit a crystalline structure akin to the cubic MgAgAs-type, featuring the

$F\bar{4}3m$ space group (Figure 1a). This structure comprisesthree interpenetrating ordered face-centered cubic (fcc) sublattices, each accommodating X, Y, and Z atoms. The corresponding occupied Wyckoff positions include 4a (0, 0, 0), 4c (1/4, 1/4,1/4), and 4b (1/2, 1/2, 1/2), while 4d (3/4, 3/4, 3/4) remains vacant. This arrangement defines the unique structural characteristics of HH compounds[9].

Diverse band engineering strategies, including resonant state doping[10], band convergence[11], and quantum confinement[12], have been implemented to finely adjust the electrical transport properties, thereby enhancing the power factor ($S^2\sigma$). Concurrently, structural manipulations such as grain refinement[13], introduction of precipitates[11], and porosity design [14] have been employed to reduce lattice thermal conductivity through reinforced phonon scattering[1]. Traditional trial-and-error experimental methods are arduous, consuming both time and resources. Moreover, the vast array of possible chemical element combinations results in exponential growth, rendering traditional labor-intensive analysesimpractical for optimizing specific materials. Theoretical calculations, employing methods like density functional theory (DFT)[15], molecular dynamics (MD) simulations[16], and nonequilibrium Green's function (NEGF) theory[17], offer reliable predictions for thermoelectric properties and insights into transport mechanisms. However, computational simulations, originally designed for simpler material systems, become computationally burdensome for complex systems involving defects, doping, and solid solutions. The high computational costs are particularly evident in high-throughput calculations.

As the demand for energy-efficient technologies grows, the development of accurate prediction models becomes paramount. In this context, data-driven methods have emerged as powerfutools in material science, facilitating the exploration of vast material spaces and accelerating the discovery of novel materials with desired properties. Contemporary research explores machine learning (ML) techniques as a viable alternative to density functional theory (DFT) calculations, delivering similarly accurate results with significantly reduced computational time and costs. These ML methods also unveil previously undiscovered correlations among seemingly unrelated material descriptors, fostering an accelerated pace in discovering and developing novel materials[18,19]. Supervised learning is prevalent in materials science, requiring sufficient relevant data with known target properties. This method has been extensively applied in the development of thermoelectric (TE) materials, encompassing predictions of the S[20], $S^2\sigma$[21], k[22] , and ZT[23] values. In machine learning (ML) applications for thermoelectric materials, the interpretation of models is crucial for understanding their effectiveness. Feature importance evaluation, a widely adopted approach, allows materials scientists to gain insights into the significance of various descriptors[24]. Notable work by Furmanchuk et al. demonstrated the importance of thermal conductivity, derived from chemical elements, in predicting S across different temperatures[20]. These research focuses on expanding the understanding and predictive capabilities for crucial thermoelectric properties— k , S, σ, and ZT—in the context of HH alloys. These materials have garnered significant attention for their promising thermoelectric performance. The approach exploits advanced ML techniques, specifically a stacking model integrating Random Forest and XGBoost, to predict these properties accurately. The significance of such predictions lies in their potential to guide the design and discovery of highly efficient thermoelectric materials. By identifying influential features and employing ensemble modelling, the study aims to contribute valuable insights to the field, addressing the complex interplay of composition, structure, and temperature in the thermoelectric behavior of HH alloys.

**computational details**

First-principles electronic structure calculations for ErNiBi and YptBi were conducted using the Quantum ESPRESSO (QE) [25] simulation package. The calculations were performed within the Generalized Gradient Approximation (GGA) employing the UltraSoft Pseudopotential type scalar relativistic pseudopotential with the Perdew–Burke–Ernzerhof (PBE) exchange–correlation functional. The undistorted structure underwent relaxation calculations to determine atomic positions near equilibrium, with forces on each atom below $10^{-3}$ eV/Å. A plane wave cut-off of 45 Ry units and a charge density cut-off of 360 Ry units were set for the plane-wave basis set on a 12 ×

12 × 12 Monkhorst–Pack K-mesh grid. Phonon dispersion and thermal properties calculations were performed using the PHONOPY [26] package interfaced with QE. PHONOPY calculates phonons at harmonic and quasi-harmonic levels, considering small atomic displacements at constant volume of the supercell. The lattice thermal conductivity was determined using the PHONO3PY [27] software package interfaced with QE, employing Q-point sampling meshes of 20 × 20 × 20 for thermal conductivity calculations and 32 × 32 × 32 for phonon dispersion. The Seebeck coeffcient, electrical conductivity and electrical thermal conductivity are calculated using the BoltzTraP2 [28] code.

**Results and discussion**

**Dataset**

The dataset, drawn from experimental results documented in various research papers, lays the foundation for an in-depth exploration of HH alloys in the realm of thermoelectric properties.

The HH dataset of several unique materials prepared by various research groups using different unique elements is shown in Figure 1b-d. This diverse collection includes 452 entries for thermal conductivity, providing a comprehensive understanding of heat transport phenomena. With 627 entries dedicated to ZT, the dimensionless figure of merit, the dataset enables precise predictions, shedding light on the overall thermoelectric efficiency of these alloys. The inclusion of 489 entries for the S spans a wide range of alloy compositions, enhancing the model's accuracy in predicting this crucial property. Simultaneously, the dataset of 444 entries for s contributes to a robust understanding of the impact of different compositions on electrical performance. The unification of datasets derived from experimental findings empowers the stacking model to make accurate predictions, thereby showcasing its potential to advance the understanding and design of thermoelectric materials.

Figure 1

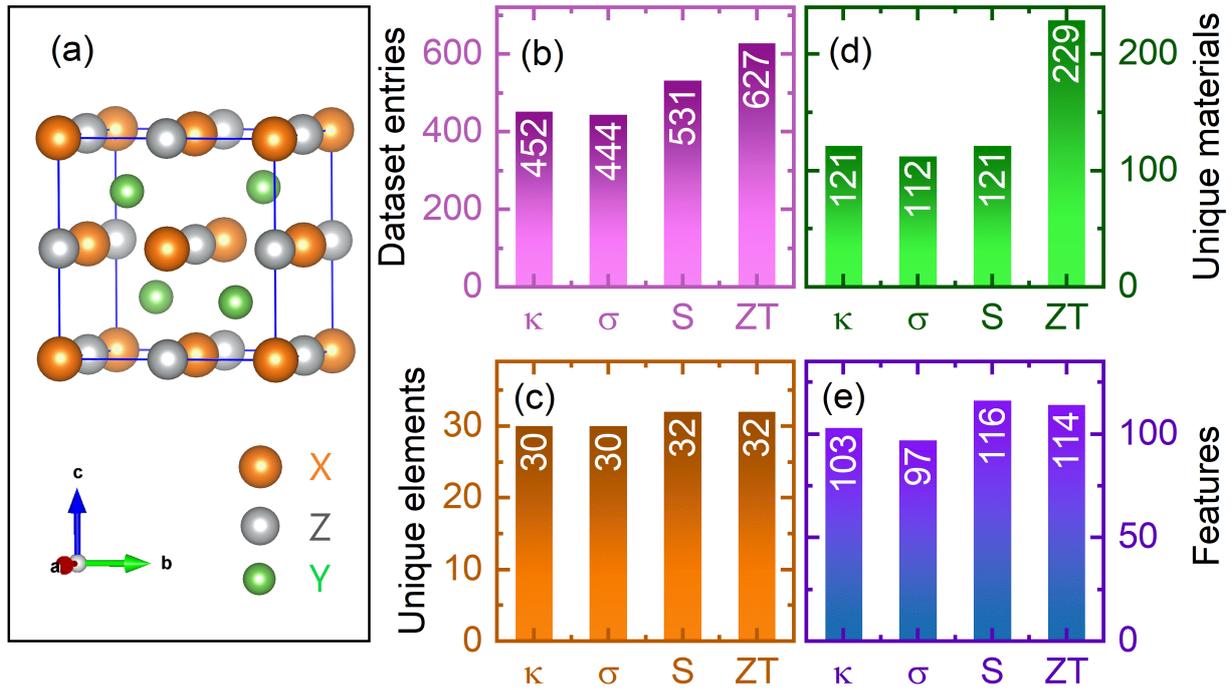

Figure 1 : (a) Schematic of crystal structure of half-Heusler alloy of class XYZ. Dataset entries (b) with available unique materials (c), unique elements (d) and features (e) for thermoelectric properties k , s, S and ZT of the half-Heusler alloys.

**Feature Generation**

The feature generation process constitutes a pivotal stage in the study, involving the extraction of compositions from chemical formulas. The 'Pymatgen' library [29] facilitated the extraction of compositions from chemical formulas and subsequent dataset transformation using the Matminer toolkit[30]. This transformation encompasses the calculation of diverse composition-based features such as stoichiometry, elemental properties, valence orbitals, and elemental fractions. The temperature was added as an additional feature and integrated into the feature matrix so that temperature-dependent thermoelectric properties can be predicted. The resultant feature matrix seamlessly integrates temperature information for a comprehensive dataset representation (Figure 1e). To enhance model efficiency, features exhibiting zero standard deviation are systematically removed. This systematic and thorough approach improves the interpretability of the ML models and equips them to predict thermoelectric properties with precision, using the distinctive compositional and temperature-related features. The quality of the generated features significantly contributes to the overall success of the stacking model in accurately predicting properties for HH alloys, showcasing theeffectiveness of the feature engineering strategy. In optimizing the input data for the ML models, a strategic step involves the removal of correlated features. The process begins by calculating the correlation matrix of the generated features, providing an overview of inter-feature relationships. Identified correlations exceeding a threshold of 0.95 are considered, and the corresponding features are excluded to mitigate redundancy. The resulting data-frame signifies the dataset after the removal of both constant and correlated columns. This process is pivotal for reducing dimensionality and enhancing the efficiency of the predictive models. Before and after removal, the number of columns and the updated feature labels are reported, offering insights into the impact of this feature selection technique. This meticulous feature engineering step contributes to refining the dataset, enabling more accurate predictions of thermoelectric properties for HH alloys in the subsequent analyses.

**Feature importance**

Feature importance evaluation also referred to as permutation feature importance, was initially introduced by Breiman [31] and has since become a valuable tool in assessing the significance of descriptors (features). This method gauges the importance of a descriptor by quantifying the rise in prediction error resulting from permuting the descriptor. Originally employed in Random Forest models[31], the permutation feature importance has evolved, and there are now model-agnostic methods available for evaluating the importance of descriptors[32]. This approach provides a robust means to identify and rank descriptors based on their impact on prediction accuracy across various machine learning models.

**Train and test set**

We utilised the shuffle function from the sci-kit-learn library to maintain data integrity and mitigate potential biases in model training[33]. This function randomly rearranges data instances, addressing any ordering effects and ensuring the exposure of the ML model to a diverse sample range. Following data shuffling, we partitioned the dataset into training and testing sets, employing a 90:10 train-test split ratio. This allocation reserved 90% of the data for model training and retained 10% for unbiased model evaluation. This division strategy allows substantial training on a diverse dataset while maintaining an independent subset for robust model assessment. To ensure feature uniformity and comparability, we implemented feature normalization. This crucial pre-processing step standardized feature scales, preventing larger-magnitude features from dominating the learning process. Mean and standard deviation calculations were performed on the training feature values, and a standardization formula was applied to both the training and testing sets. Specifically, each feature value underwent subtraction by its mean and division by its standard deviation. These standardized features facilitate fair comparisons and accurate model training.

**ML Model**

Stacking, or stacked generalization, is an ensemble learning technique where multiple base models are trained to make predictions, and a meta-model is then trained to combine these predictions[34,35]. The stacking technique enhances the model performance by combining diverse yet effective models, aiming to reduce the generalization error. In our approach, two ensemble learning models, namely Random Forest[36], and Xgboost[37], constituted the first layer of the stacking model. The featured support vector regression (SVR)[38], the meta-model, forms the second layer. The stacking model architecture, depicted in Figure 2, involved training the initial base models using 5-fold cross-validation. Subsequently, predictions from these base models served as inputs for the meta-model. A GridSearchCV method with 5-fold cross-validation and a comprehensive grid search were employed to optimize model hyperparameters. Model performance was evaluated using the coefficient of determination ($R^2$), providing a robust metric for assessing the efficacy of the ML models. The advantage of a stacking model lies in its ability to harness the strengths of diverse base models, mitigating their weaknesses. The diverse algorithms capture different aspects and patterns within the data, and the meta-model learns how to best combine their predictions. This ensemble approach often leads to improved predictive performance compared to individual models.

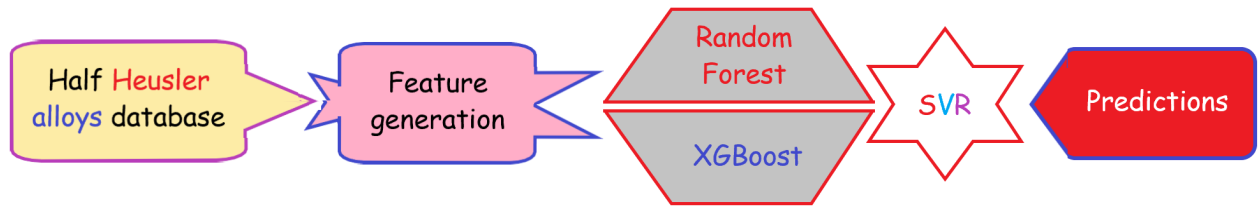

*Figure 2 : Schematic of the stacking model architecture using the Random Forest and XGBoost as the base models.*

**Predictions of stacking model on different thermoelectric properties**

**1. Thermal conductivity ( k )**

Thermal conductivity, a fundamental physical quantity, characterizes heat transfer within a material when subjected to thermal energy. Its significance extends beyond fundamental physics and is crucial in practical applications like thermal management for thermoelectric energy conversion devices and spintronics technology[39]. The separation of thermal conductivity contribution from the electron and lattice provides insights into material behavior. HH compounds, characterized by high crystal symmetry, the lattice thermal conductivity is notably elevated, emphasizing the need for precise knowledge in thermal management for devices based on these materials. Understanding and controlling thermal conductivity is essential for optimizing the performance, lifespan, and safety of such technologies. Studies have shown that predicting the lattice thermal conductivity of HH compounds within a 10% range is feasible, and their approach utilizes Young's modulus value obtained from density functional theory (DFT) calculations as a descriptor for ML[23]. Another method involves using descriptors such as atomic numbers, atomic masses, and atomic radii of constituent atoms, achieving a comparable predictive accuracy for lattice thermal conductivity in HH compounds[40].

Figure 3

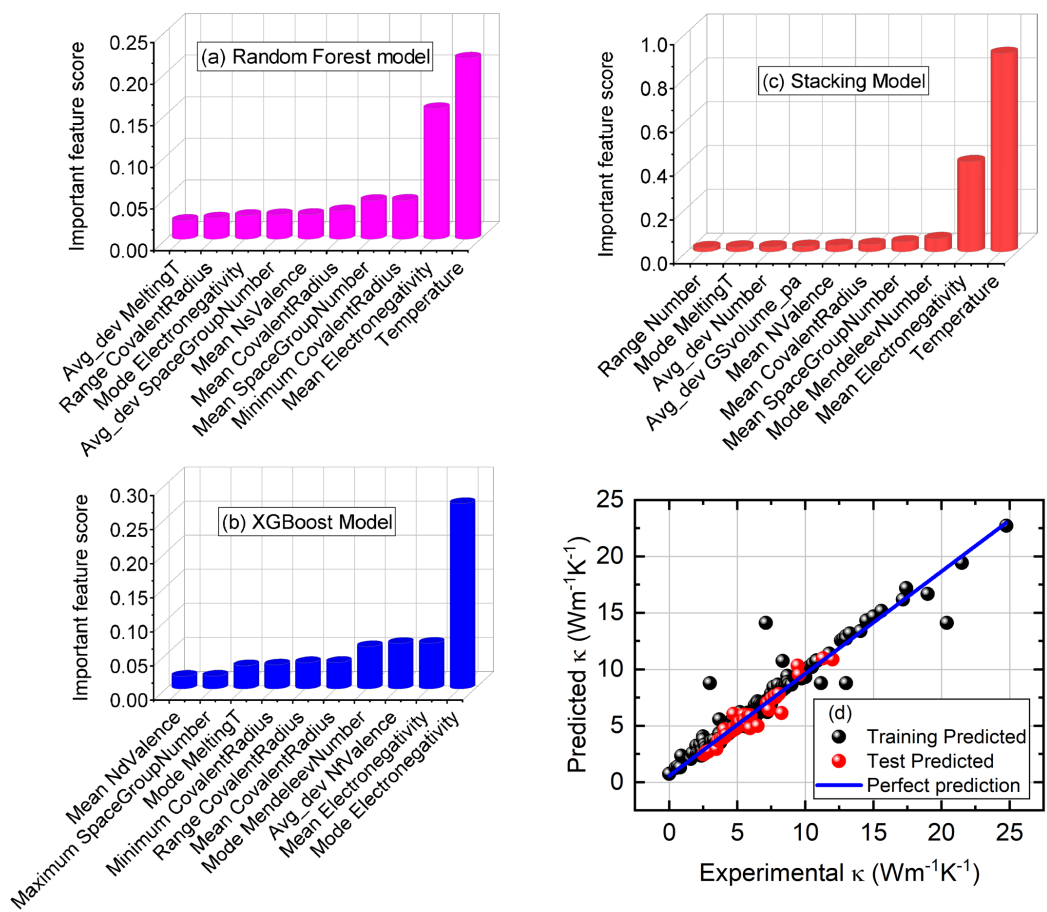

*Figure 3: Important features and corresponding scores of thermal conductivity in the (a) Random Forest, (b) XGBoost, (c) stacking models. (d) The predictions of thermal conductivity using the stacking model.*

The ensemble, combining Random Forest and XGBoost, demonstrates remarkable efficacy in predicting lattice thermal conductivity (Figure 3). The foremost feature, "Temperature," emerges as a pivotal factor, highlighting the inherent sensitivity of thermal conductivity to temperature variations. The well-established temperature-dependent nature of thermal conductivity underscores its crucial role in governing heat transport in HH alloys. Additionally, "Mean Electronegativity" proves to be a significant feature, encapsulating the average electronegativity of constituent elements and emphasizing the influence of electronic properties on thermal transport. In the Random Forest model, descriptors like "Minimum CovalentRadius" and "Mean SpaceGroupNumber" are underscored, providing insights into the structural characteristics of the material (Figure 3a). These features offer information on atomic sizes and crystal symmetry, impacting lattice thermal conductivity. Meanwhile, the XGBoost model highlights "Mode Electronegativity" and "Avg_dev NfValence," pointing towards the role of electronegativity and variations in the number of valence electrons in influencing thermal conductivity (Figure 3b). The consistent presence of "Temperature" and "Mean Electronegativity" across all models reaffirms their universal significance (Figure 3c). The intricate interplay of these features, rooted in the intrinsic properties of the materials, contributes to the understanding of thermal transport mechanisms. For instance, temperature- dependent behavior reflects how heat is conducted through the lattice structure, while electronegativity-driven characteristics shed light on the impact of electronic properties. This comprehensive analysis provides valuable insights into the underlying factors influencing thermal conductivity and informs material design strategies for optimizing thermoelectric applications. The model performance, assessed by the coefficient of determination $R^2$, underscores the superiority of the stacking model over individual base models, boasting an impressive $R^2$ of 0.93 (Table -1). The training and test predicted k of HH compounds is compared with experimental k in Figure 3d. The solid line represents an ideal plot for a perfect match when the experimental and predicted data are exactly equal. The comparison of the experimental and predicted k demonstrates the accuracy of the stacking model with the $R^2$ value of 0.93

| parameters | Random Forest | XGBoost | Stacking model |
| --- | --- | --- | --- |
| Thermal Conductivity | 0.22 | 0.92 | 0.93 |
| Electrical Conductivity | 0.94 | 0.96 | 0.96 |
| Seebeck coefficent | 0.88 | 0.97 | 0.99 |
| Figure of Merit (ZT) | 0.32 | 0.91 | 0.91 |

*Table – 1. Comparison of $R^2$ values of stacking model with the base models.*

**2. Electrical conductivity ( σ )**

The σ, a fundamental property in thermoelectricity, measures a material's capacity to conduct electric current. In thermoelectric devices, efficient charge carrier movement is crucial for optimal performance. In a study by Mukherjee et al., an ML approach was introduced for the rapid prediction of σ. The method employed a gradient boosting regression (GBR) model, initially trained on 124 experimental data points, to predict log-scaled conductivity[41]. This approach aimed to leverage the power of ML for efficient and accurate estimations of σ in materials. In our rigorous examination of σ in HH alloys, the R-squared ($R^2$) values serve as robust performance benchmarks, with Random Forest achieving an $R^2$ of 0.94, XGBoost exhibiting a noteworthy $R^2$ of 0.96, and the stacking model excelling with a matching $R^2$ of 0.96 (Table -1). A key universal influencer, "Temperature," consistently emerges as a predominant factor across all models, emphasizing its central role in governing electrical conductivity (Figure 4). In the stacking model, the "Avg_dev Column" provides information about the average deviation of the periodic table column numbers for the elements in a given material composition and points towards insights into the electronic structure and composition of materials, which are crucial factors influencing σ. Simultaneously, "Avg_dev Gsvolume_pa" holds importance, indicating the influence of crystal volume per atom on the electrical conducting properties of HH. The emphasis on these features underscores the importance of structural considerations in predicting σ of HH. Similarly, the Random Forest Model prioritizes "Avg_dev Column" and "Avg_dev GSvolume_pa," emphasizing their relevance to σ and k. The attention given to structural variations and crystal volume in both ensemble and individual models highlights their substantial impact on the occurrence σ in HH. Crucially, "Mean Electronegativity" emerges as a pivotal feature in both the Stacking and XGBoost models, highlighting the role of mean electronegativity in predicting s. Electronegativity, representing the tendency of an atom to attract electrons, plays a significant role in electron conduction

Figure 4

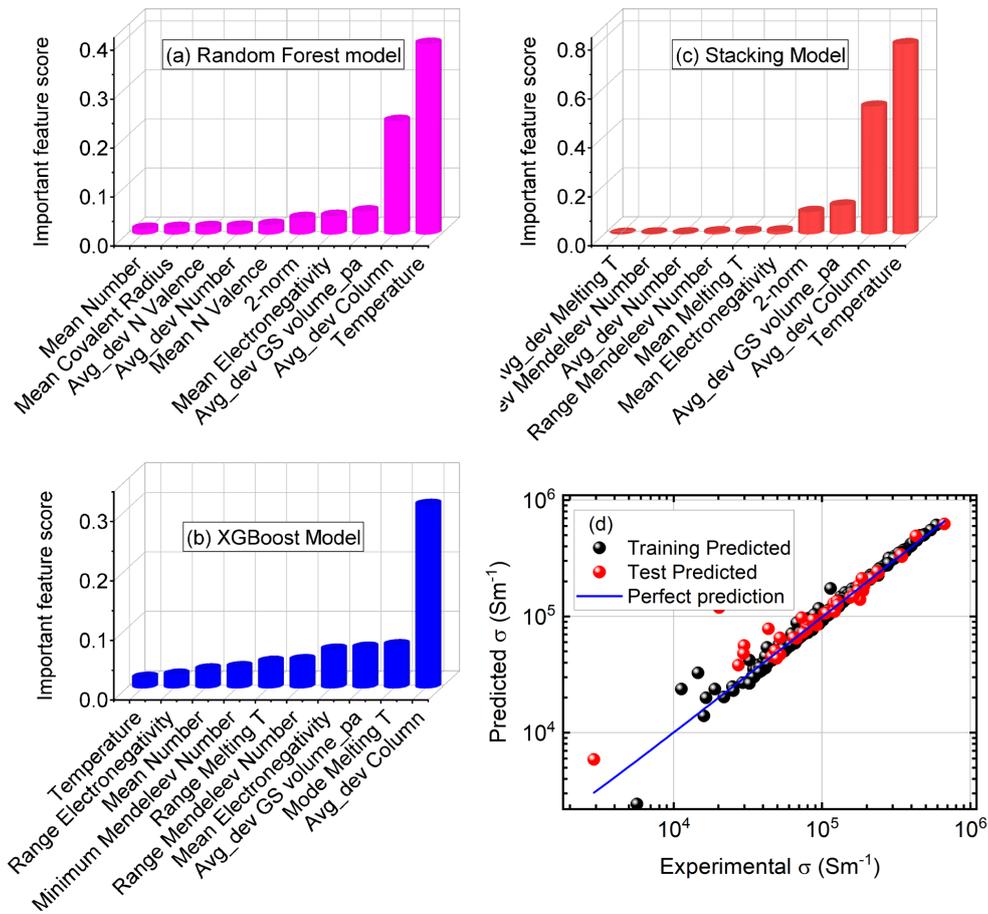

*Figure 4 : Important features and corresponding scores of conductivity in the (a) Random Forest, (b) XGBoost, (c) stacking models. (d) The predictions of electrical conductivity using the stacking model.*

**3. Seebeck coefficient (S)**

The S holds significant importance in thermoelectric properties, as it exhibits a dominant role in the thermoelectric figure of merit (ZT $\propto$ $S^2$). This correlation is crucial as it directly influences the efficiency of energy conversion processes in thermoelectric materials[42]. The S is defined as the ratio of the voltage difference applied to a thermoelectric material to the temperature gradient. It quantitatively characterizes the material's capability to convert heat directly into electricity. While experimental measurements or theoretical calculations can provide the S of a material, these processes often entail significant costs and efforts. ML has emerged as a prominent research focus in the thermoelectric field, offering efficient processing of numerous candidate materials. Notably, Furmanchuk et al. introduced the Random Forest regression algorithm to predict the S across various temperatures (300, 400, 700, and 1000 K). The model utilized input features such as experimental synthesis conditions, properties of constituent elements, and compound crystallinity, leveraging data from S measurements for 130 materials[20]. In another study, Yuan et al. harnessed 856 n-type and 909 p-type S entries to construct an Artificial Neural Network (ANN) model. This model was designed to predict the S of Heusler compounds across various carrier concentrations, specifically at room temperature[43].

In the present study, the stacking model excelled with an impressive $R^2$ of 0.99, underscoring the efficacy of the model ensemble (Table 1). In the analysis of feature importance across the Stacking, Random Forest, and XGBoost models, several common features emerged as influential contributors to the prediction of the S. Notably, the temperature feature exhibited substantial importance across all models, emphasizing its fundamental role in governing the thermoelectric behavior of HH materials. The "Mean CovalentRadius" and "Avg_dev MendeleevNumber" also demonstrated consistent importance, suggesting their significant impact on the S. These features likely capture intricate relationships related to the geometric and structural aspects of materials, influencing charge carrier mobility and, consequently, the S. Moreover, features such as "Mean MeltingT" and "Minimum Electronegativity" in multiple models imply a connection between thermal stability and electronic properties, offering insights into how temperature-dependent phenomena influence the S. The "2-norm" and "Mean NpUnfilled", unique to the Stacking and XGBoost models, respectively, contribute nuanced information regarding the stoichiometric and electronic structure of materials (Figure 5). The experimental and stacking model predicted S is plotted in Figure 5d. Overall, Figure 5d shows the negligible number of overestimated and underestimated predicted S with $R^2$ score of

0.99. This comprehensive analysis provides valuable guidance for tailoring material compositions to optimize the S, a critical parameter in enhancing the efficiency of thermoelectric devices.

Figure 5

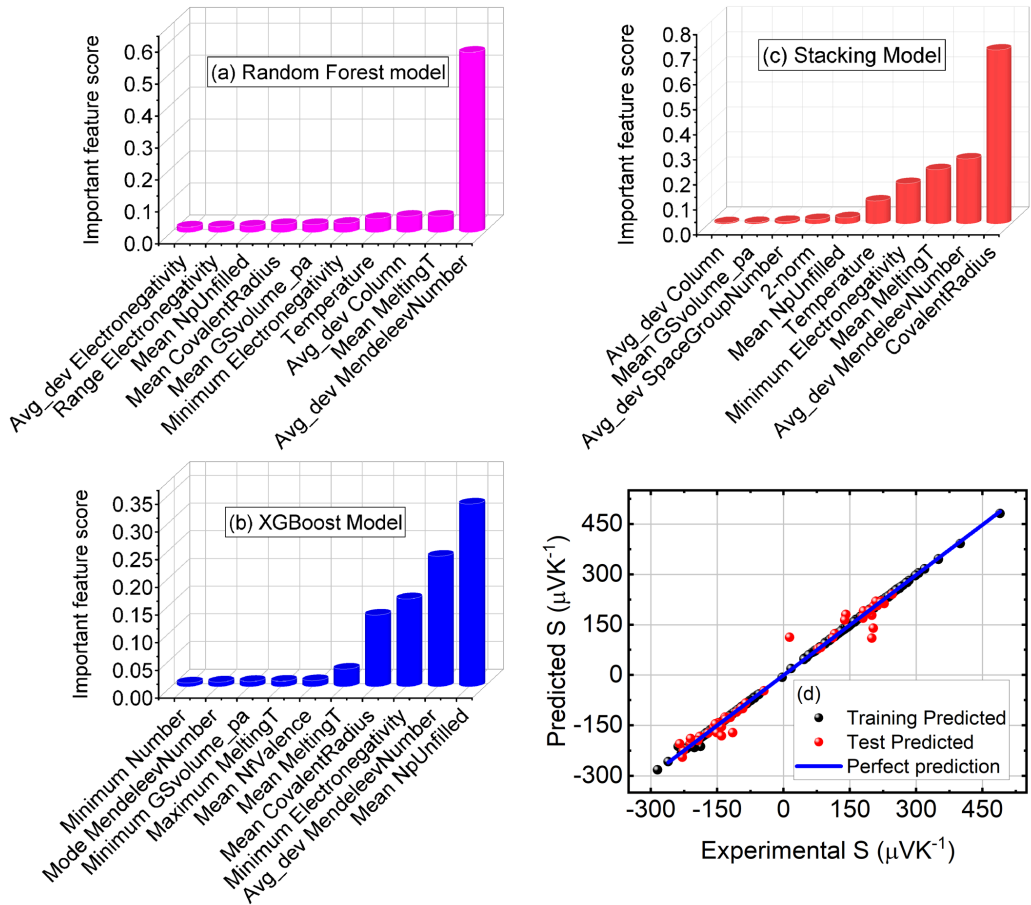

*Figure 5 : Important features and corresponding scores of Seeback coefficient in the (a) Random Forest, (b) XGBoost, (c) stacking models. (d) The predictions of Seeback coefficient using the stacking model.*

## 4. Figure of merit(ZT)

The ZT is a crucial parameter for accessing the suitability of a thermoelectric material, representing the efficiency in converting heat energy into electrical energy. It quantifies the ability of a material to generate a voltage when subjected to a temperature gradient. A higher ZT value indicates a more efficient thermoelectric material, capable of achieving greater energy conversion efficiency. Researchers aim to enhance ZT by optimizing the material's electronic and thermal transport properties. Therefore, ZT serves as a crucial metric for evaluating and comparing the performance of thermoelectric materials, guiding the design and development of materials for applications such as waste heat recovery and energy harvesting.

In our pursuit of precise ZT prediction for HH alloys, a stacking model was employed alongside Random Forest and XGBoost as base models. The $R^2$ values served as performance metrics, indicating the effectiveness of each model: Random Forest achieved an $R^2$ of 0.32, XGBoost demonstrated notable performance with an $R^2$ of 0.91, and the stacking model excelled with an impressive $R^2$ of 0.92, showcasing the strength of model ensemble (Table-1).

The ZT evaluation for thermoelectric materials involves common features identified by the stacking model, Random Forest, and XGBoost models. "Temperature" is consistently recognized as a crucial factor across all models, emphasizing its fundamental role in influencing ZT. Moreover, the minimum electronegativity is highlighted in the stacking and RF models, indicating its significant impact on ZT. This feature reflects the electronic properties of the material, contributing to improved energy conversion efficiency. In the XGBoost model, other vital features influencing ZT include "Minimum electronegativity", "Maximum Gsmagmom" (insights into magnetic behavior), and "Mode SpaceGroupNumber", showcasing the intricate interplay between electronic and structural characteristics (Figure 6). The critical analysis of these features underscores their collective contribution to optimizing ZT, offering valuable insights for the strategic design of thermoelectric materials for enhanced performance. In the exploration of HH alloys, the stacking model, integrating with Random Forest and XGBoost, has exhibited remarkable predictive prowess, surpassing individual base models with high $R^2$ values of 0.93 for thermal conductivity, 0.96 for electrical conductivity, 0.99 for the Seebeck coefficient, and 0.92 for ZT. Crucial features, namely "Temperature", "Mean CovalentRadius", and "Avg_dev GSvolume_pa", emerge as linchpins in forecasting these thermoelectric properties, signifying their intricate roles in thermoelectricity optimization.

Figure 6

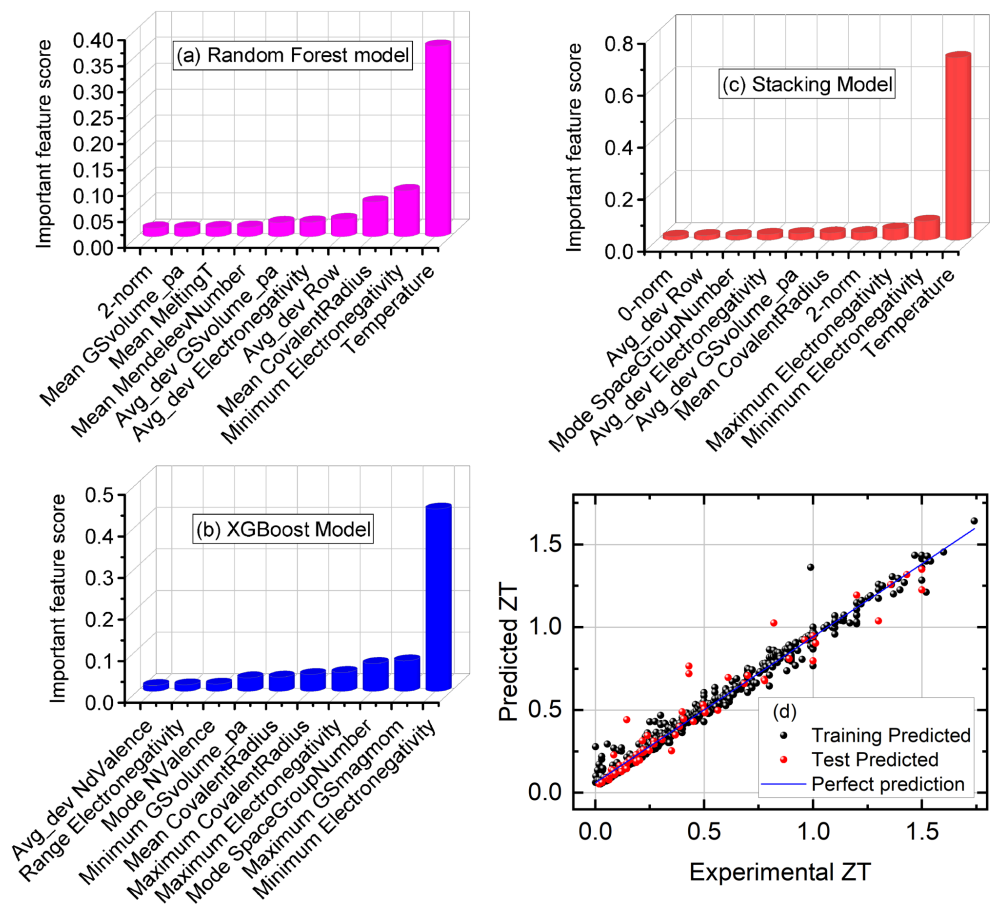

*Figure 6 : Important features and corresponding scores of the figure of merit in the (a) Random Forest, (b) XGBoost, (c) stacking models. (d) The predictions of the figure of merit using the stacking model.*

As a primary factor, the "Temperature" influences k, s, and the σ. Its impact on the S, which gauges a material's capacity to convert heat into electricity, is particularly noteworthy. The "Mean CovalentRadius", indicative of the average atomic size, holds sway over the material's structural and electronic attributes, influencing charge carrier mobility and electrical conductivity. Additionally, "Avg_dev GSvolume_pa", representing the material's stability and energy landscape, assumes significance in the thermoelectric context. This descriptor sheds light on how the material responds to temperature variations, providing insights into its thermal transport properties. The intricate interplay of these features manifests in optimized thermoelectric performance. For instance, a higher "Mean CovalentRadius" enhances structural stability, reducing phonon scattering and bolstering electrical conductivity. Simultaneously, a stable energy landscape contributes to efficient charge carrier transport. The stacking model's amalgamation of Random Forest and XGBoost effectively captures these nuanced relationships, offering a holistic comprehension of HH alloys' thermoelectric behavior. The obtained high $R^2$ values underscore the efficacy of this ensemble modelling approach. This comprehensive understanding of the connected influence of "Temperature", "Mean CovalentRadius", and "Avg_dev GSvolume_pa" on stacking model electric properties is pivotal for the strategic design of materials, opening avenues for advancements in stacking model electric applications with enhanced efficiency and performance.

The performance of the stacking model in predicting the ZT is shown in Figure 6d. Figure 6d includes the training and test predicted ZT for comparison with the experimental ZT and the perfect prediction line. The presence of very few overestimated and underestimated predictions in the ZT range of 0.0017 to 1.74 with the $R^2$ value of 0.92 demonstrates unparalleled predictive accuracy, and outperforming the individual contributions of Random Forest and XGBoost.

**Conclusion**

In conclusion, the stacking model electric properties of HH alloys were explored thoroughly by leveraging a robust dataset with a sophisticated ensemble modelling approach. Specifically, the stacking model integrates Random Forest and XGBoost. The predictive performance of the stacking model surpassed individual base models, achieving impressive $R^2$ values across critical stacking model electric properties—0.93 for k , 0.99 for the S, 0.96 for σ, and 0.92 for ZT. The stacking model demonstrated unparalleled predictive accuracy, outperforming the individual contributions of Random Forest and XGBoost across all properties. Noteworthy features, such as "Temperature", "Mean CovalentRadius", and "Avg_dev Gsvolume_pa", consistently emerged as influential factors, unravelling the intricate relationships dictating stacking model electric behavior in these alloys. The stacking model's versatility, coupled with the identification of common influential features, establishes a robust framework for simultaneously predicting multiple stacking model electric properties. This work advances our understanding of the fundamental factors steering stacking model electric performance in HH alloys, and it furnishes materials scientists with a powerful tool to tailor compositions for heightened stacking model electric efficiency. The synergistic combination of ensemble modelling and feature insights sets the stage for accelerated strides in the design and discovery of high-performance stacking model electric materials.

AUTHOR INFORMATION:

Corresponding Author: ph18d200@smail.iitm.ac.in

CONFLICT OF INTEREST

The authors declare no competing interest.

DATA AVAILABILITY

The data that support the findings of this study are available from the corresponding authors upon reasonable request.

**Acknowledgements:**

We would like to acknowledge the use of the computing resources at High Performance Computing Environment, Indian Institute of Technology Madras.